\documentclass[twocolumn,amssymb,superscriptaddress,showpacs,aps,prl]{revtex4-1}
\usepackage{}
\usepackage{stmaryrd}
\usepackage{amsmath}
\usepackage{amssymb}
\usepackage{float}
\usepackage{array, color}
\usepackage{tabularx}
\usepackage{threeparttable}
\usepackage{titletoc}
\usepackage{booktabs}
\usepackage[colorlinks,linkcolor=blue,anchorcolor=blue,citecolor=blue,urlcolor=blue,dvipdfm]{hyperref}
\usepackage{graphicx}
\usepackage{subfigure}
\usepackage{latexsym,bm}
\usepackage{graphics}
\newcommand{\ie}{i.\,e.}

\begin{document}

\title{Revealing the Complex Transport Behaviors in Warm Dense Hydrogen by Including Nuclear Quantum Effects}

\author{Dongdong Kang}
\author{Huayang Sun}
\author{Jiayu Dai}
\email{jydai@nudt.edu.cn}
\author{Zengxiu Zhao}
\author{Yong Hou}
\author{Jiaolong Zeng}
\affiliation{Department of Physics, College of Science, National University of Defense Technology, Changsha 410073, Hunan, People's Republic of China}
\author{Jianmin Yuan}
\email{jmyuan@nudt.edu.cn}
\affiliation{Department of Physics, College of Science, National University of Defense Technology, Changsha 410073, Hunan, People's Republic of China}
\affiliation{State Key Laboratory of High Performance Computing, National University of Defense Technology, Changsha 410073, Hunan, People's Republic of China}
\date{\today}

\begin{abstract}

Nuclear quantum effects (NQEs) on the structures and transport
properties of dense liquid hydrogen at densities of 10$\sim$100 g/cm$^3$ and temperatures of 0.1$\sim$1 eV are fully assessed using \textit{ab initio} path-integral molecular dynamics simulations. With the inclusion of NQEs, ionic diffusions are strongly enhanced by the magnitude from 100\% to 15\% with increasing temperature, while electrical conductivities are significantly suppressed. The analyses of ionic structures and zero-point energy show also the importance of NQEs in these
regime. The significant quantum delocalization of ions introduces expressively different scattering cross section between protons compared with classical particle treatments, which can explain the large alterability of transport behaviors. Furthermore, the energy, pressure, and isotope effects are also greatly influenced by NQEs. The complex behaviors show that NQEs can not be neglected for dense hydrogen even in the warm dense regime.
\end{abstract}

\pacs{62.50.-p, 66.10.-x, 72.15.Cz, 67.90.+z}

\maketitle

Dense liquid hydrogen plays a crucial role in understanding the material behaviors under extreme conditions \cite{Fortov,McMahon1,Babaev} and many applications in astrophysics and energy sources \cite{Lambert,Glenzer,Guillot,Fortney,Militzer}. Ionic and electronic transport properties of warm and hot dense hydrogen are key points in the dynamics of capsule implosion of inertial confinement fusion (ICF) \cite{Lambert}, modeling plasma processes \cite{Daligault} and the interior structure and evolution of giant planets and exoplanets \cite{Guillot,Fortney,Santos}, of which the core pressures are estimated even up to 19 Gbar \cite{Swift}, while the temperature may be down to thousands of Kelvin. Small change of transport properties will remarkably alter the dynamics of planet's evolution \cite{AlfeNat} and hydrodynamical processes in ICF \cite{Hu}. Thanks to the recent progress in experimental techniques, one could get access to this previously inaccessible regime of phase diagram of hydrogen through a laser-induced shock wave loading of precompressed samples \cite{Loubeyre,Jeanloz}. However, theoretically understanding the nature of hydrogen under the ultra-high pressures is still rare and a great challenge. Density functional theory based molecular dynamics simulations, named quantum molecular dynamics or \textit{ab initio} molecular dynamics (AIMD) is one of the most successful method for (warm) dense matter \cite{Collins}, but typically employ a classical-particle approximation for ions. Using AIMD, transport properties at high temperature such as ionic diffusion, electrical conductivity have been obtained with good accuracy \cite{Alfe,Holst,Lambert2}. Unfortunately, this approximation could lead to inaccurate results, especially when the nuclear quantum character is noticeable for light-atom systems \cite{Benoit,Morrone,Herrero1,Li,Witt}.

The protons in condensed hydrogen easily exhibit quantum effects due to their low masses. These quantum effects include nuclear quantum fluctuations and exchange effects involving Bose-Einstein or Fermi-Dirac statistics \cite{Dyugaev}. Path-integral molecular dynamics (PIMD) \cite{Feynman,Marx1,Tuckerman,Ceperley} have provided a proper description of the nuclear quantum fluctuations beyond the harmonic level. Recently, a lot of simulations largely focus on the role of nuclear quantum effects (NQEs) in the structures of dense hydrogen below several 1000 K, showing rich phases and dynamics \cite{Biermann,Kitamura,Geneste,Morales} induced by quantum fluctuations or zero-point motion (ZPM). Compared with the classical-particle treatment, large deviation for ionic diffusion and thermal conductivity of para-hydrogen is found below 32 K induced by NQEs \cite{Yonetani}. People usually consider ions as distinguishable classical particles in warm dense regime (\ie, above 1000 K) since the quantum fluctuations or quantum collisions are generally regarded to be suppressed at high temperatures. However, even for the collision energies up to 10 eV,
it has been shown that elastic collision cross sections for protons
can be different significantly in the quantum theory \cite{cross}
from the classical treatment. Nevertheless, NQEs are generally ignored
in \textit{ab initio} simulations of hydrogen in warm dense regime.
This classical treatment for nuclear motion would be insufficient
for dense hydrogen. In viewpoint of collision physics, both the
ionic and electronic transport properties depend strongly on the
scattering cross sections between particles. In usual AIMD, the corresponding scattering cross sections are statistically obtained from classical particle dynamics simulations and therefore are expected to differ from the results with NQEs.

In this Letter, we aim to fully assess NQEs on the structures and transport properties of warm dense hydrogen using \textit{ab initio} PIMD (AI-PIMD). The results show that the NQEs play an impressive role on both static and dynamic properties and can not be neglected for dense hydrogen even at the high temperature of 1 eV.

Electronic structure calculations were carried out based on pseudopotential with plane-wave expansion of electronic wave functions \cite{PP,Dai2} and generalized-gradient approximation for exchange and correlation \cite{PBE}. Electronic distribution is subjected to Fermi-Dirac statistics \cite{Mermin}. A periodic supercell including 250-432 atoms was employed according to different densities, which can ensure convergence of both ionic and electronic properties with good accuracy \cite{Lambert,Dai4}. A bcc lattice was used as the initial structure according to the results in Ref.~\onlinecite{Liberatore}. AI-PIMD calculations as well as their classical counterparts in this work were performed with modified Quantum-ESPRESSO package \cite{QE}. Within the framework of AI-PIMD, the structural and thermodynamical properties were calculated using the primitive scheme \cite{Marx1,Tuckerman}, while the real-time quantum dynamics of nuclei was obtained through the \textit{ab initio} centroid path-integral molecular dynamics (AI-CMD) \cite{Cao,Marx2}. The self-diffusion coefficients of ions were calculated from the slope of the mean square displacement obtained from AI-CMD simulations. Langevin thermostat \cite{Gillan} was employed to overcome the nonergodic problem, which not only produces a canonical ensemble and compensates the calculated errors \cite{Dai2}, but also gives us an efficient unified description from cold condensed matter to hot dense regime \cite{Dai2,Dai3}. The Trotter number was set to 16 after a convergence test \cite{Supplement}. The proper time step and sufficient total steps were employed \cite{Timestep}. The electrical conductivity and optical absorbtion coefficient were calculated via the Kubo-Greenwood formula \cite{kubo,green} with ABINIT package \cite{ABINIT}.

The density of hydrogen under consideration ranges from 10 g/cm$^3$ to 100 g/cm$^3$, corresponding to 0.3 $\leq r_s \leq$ 0.65 ($r_s$ denotes the Wigner-Seitz sphere radius). For each state in our simulations, the temperature ranging from 0.1 eV to 1 eV is well above the corresponding quantum degeneracy temperature \cite{Dyugaev,Bermejo}, thus the exchange of particles is negligible.

\begin{figure}[t]
\centering
\includegraphics*[width=7.5cm]{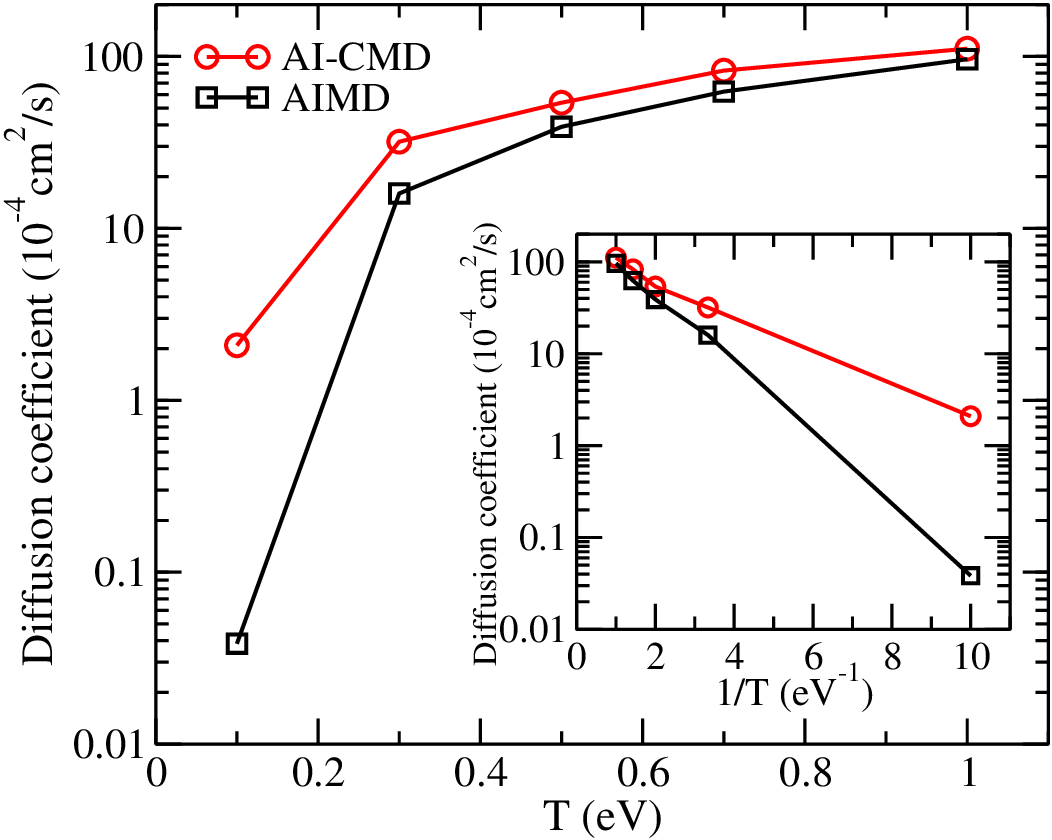}
\caption{(color online). Temperature dependence of the self-diffusion coefficients of hydrogen atoms at 10 g/cm$^3$. For comparisons, both the results from AI-CMD and AIMD simulations are presented. The inset is the Arrhenius plot of the self-diffusion coefficient.}
\label{fig1}
\end{figure}

\begin{figure}[t]
\centering
\includegraphics*[width=8cm]{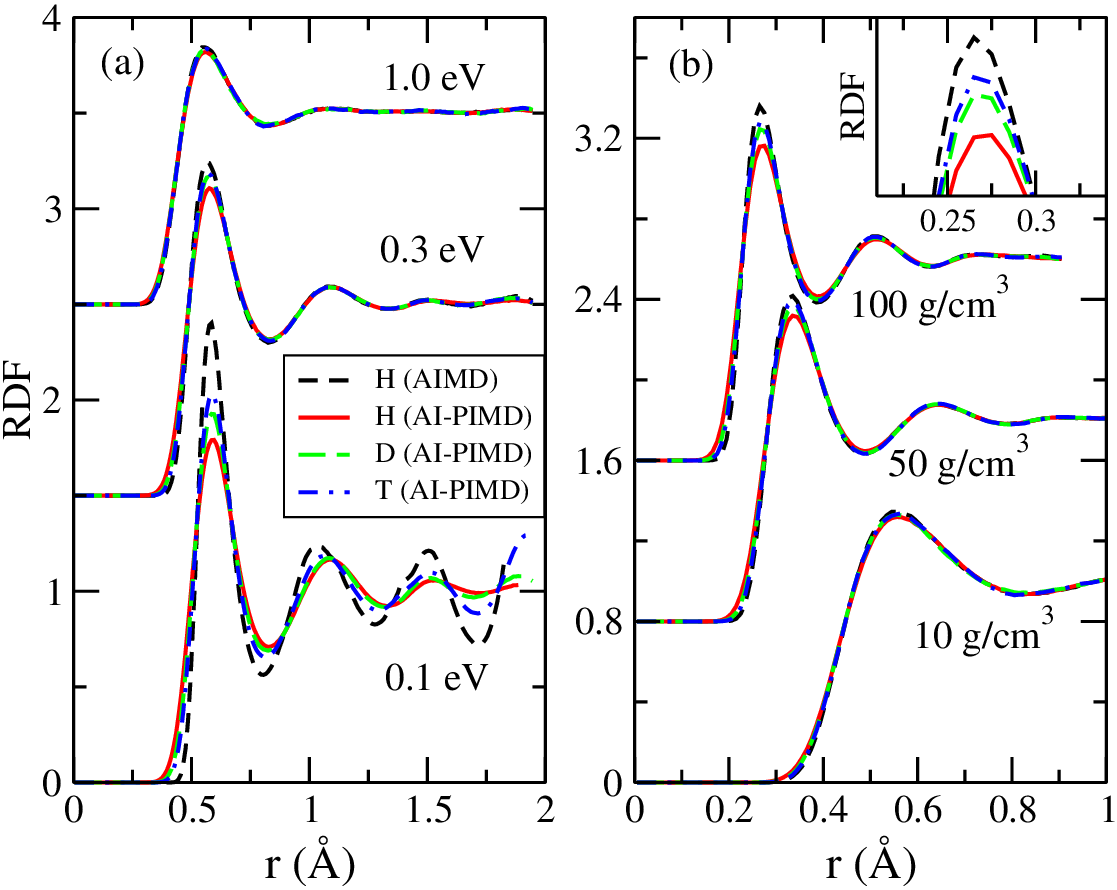}
\caption{(color online). Radial distribution functions of hydrogen nuclei from AI-PIMD (solid lines) and AIMD (dashed lines) simulations with different temperatures at 10 g/cm$^3$ (a), and with different densities at 1 eV (b). The isotopes effects of deuterium (D) (long-dashed lines) and tritium (T) (dot-dashed lines) are also presented. The inset is the first peak of radial distribution functions at 100 g/cm$^3$ and 1 eV.}
\label{fig2}
\end{figure}

Firstly, we calculated self-diffusion coefficients of hydrogen atoms at 10 g/cm$^3$ with temperatures from 0.1 eV to 1 eV, as shown in Fig.~\ref{fig1}. There are remarkable differences between the results from AIMD and AI-PIMD simulations. First of all, the self-diffusion coefficient at 0.1 eV from the AI-PIMD simulations is more than one order of magnitude larger than the AIMD value. This amazing increment is likely to arise from the phase transition from solid to liquid with the inclusion of NQEs. It is confirmed directly by ionic trajectory analyses (See the Supplemental Material for simulated movies \cite{Supplement}), where in the AI-PIMD simulations hydrogen exhibits obvious liquid character, whereas in the AIMD simulations hydrogen atoms remain bcc structure with large vibrational amplitudes. The difference introduced from NQEs is also clearly demonstrated by radial distribution functions (RDFs) of hydrogen nuclei in Fig.~\ref{fig2}(a). We note that the RDF of hydrogen nuclei with AIMD method exhibits long-range ordered solid-like character at the low temperature of 0.1 eV, which is not visible in RDF obtained from the AI-PIMD calculations. In the quantum-mechanical perspective, it can be understood from that the protons tunnel through the energy barriers and move away from their equilibrium positions in bcc structure at 0.1 eV, as appeared in hydrogen-bonded systems at low temperatures \cite{Benoit}. Our results accord with the conclusions in Ref. \onlinecite{Liberatore} that the melting temperature of hydrogen including NQEs is lowered compared with the classical particle treatment.

Interestingly, the self-diffusion coefficients obtained via AI-CMD are substantially larger than the classical particle value over the whole temperature range of 0.1-1 eV, as shown in Fig.~\ref{fig1}. The difference between them is from 100\% at 0.3 eV, 38\% at 0.5 eV to 32\% at 0.7 eV. There is still a large difference of 15\% even at the high temperature of 1 eV. It is not open-and-shut because the RDFs from the AIMD and AI-PIMD calculations are very close at 0.3 eV and almost identical when the temperature is up to 1 eV (see Fig.~\ref{fig2}(a)). In addition, it is well-known that in the regime satisfied Boltzmann statistics, the diffusion is thermally activated and the temperature dependence of the self-diffusion coefficient $D(T)$ generally obey the Arrhenius law, $D(T)\propto \exp(-E_a/k_BT)$ \cite{Dyer}, where the activation energy $E_a$ is the energy required for the hydrogen to surmount the potential barrier separating the neighboring configurations. Here, the Arrhenius plot (see the inset in Fig.~\ref{fig1}) shows that the self-diffusion coefficients from the AIMD calculations exhibit the Arrhenius behavior above 0.3 eV. The diffusion at 0.1 eV is smaller than the value according to the Arrhenius relation, arising from the solidification in the AIMD simulations as mentioned above. On the contrary, the results from the AI-CMD calculations are obviously larger than the estimates of the Arrhenius relation below 0.5 eV, indicating that quantum diffusion mechanism is becoming crucial in the temperature range. We note that the large-angle scattering between ions is dominant at such high densities considered here. The pronounced quantum nuclear wave scattering
calculation leads to a smaller large-angle scattering cross section between ions than the classical particle scattering calculation, and thus increases the mean free path of ions. Therefore, the ions diffuse more easily even beyond the quantum tunneling regime and the classical particle treatment of protons substantially underestimate the ionic diffusion. In fact, when the indistinguishable effects are notable, much more  significant effects will be visible even up to 10 eV \cite{cross}. This microscopic interpretation allows one to understand profoundly the transport properties of dense matter even at high temperatures where NQEs are conventionally ignored in the AIMD simulations.

The largely reduced and broadened first peak of RDF from the AI-PIMD calculations at 0.1 eV and 10 g/cm$^3$ indicates the significant nuclear quantum delocalization, which becomes weaker with increasing temperature and can not be observed at 1 eV finally. In contrast, RDFs show much pronounced nuclear quantum character with increasing density (see Fig.~\ref{fig2}(b)). When the density is increased up to 100 g/cm$^3$, there is a distinct difference of RDFs between the AI-PIMD and AIMD calculations due to NQEs even at the high temperature of 1 eV. Meanwhile, the isotope substitutions provide a clear demonstration of isotope effects on nuclear spacial distribution of condensed hydrogen. Since the tritium nuclei is the heaviest one and has the shortest thermal de Broglie wavelength \cite{Supplement}, RDFs of the tritons are more structured than those of the other two isotopes, and more close to the results of the AIMD simulations for protons as the experimental observations \cite{Giuliani}.

\begin{figure}[t]
\centering
\includegraphics*[width=8.5cm]{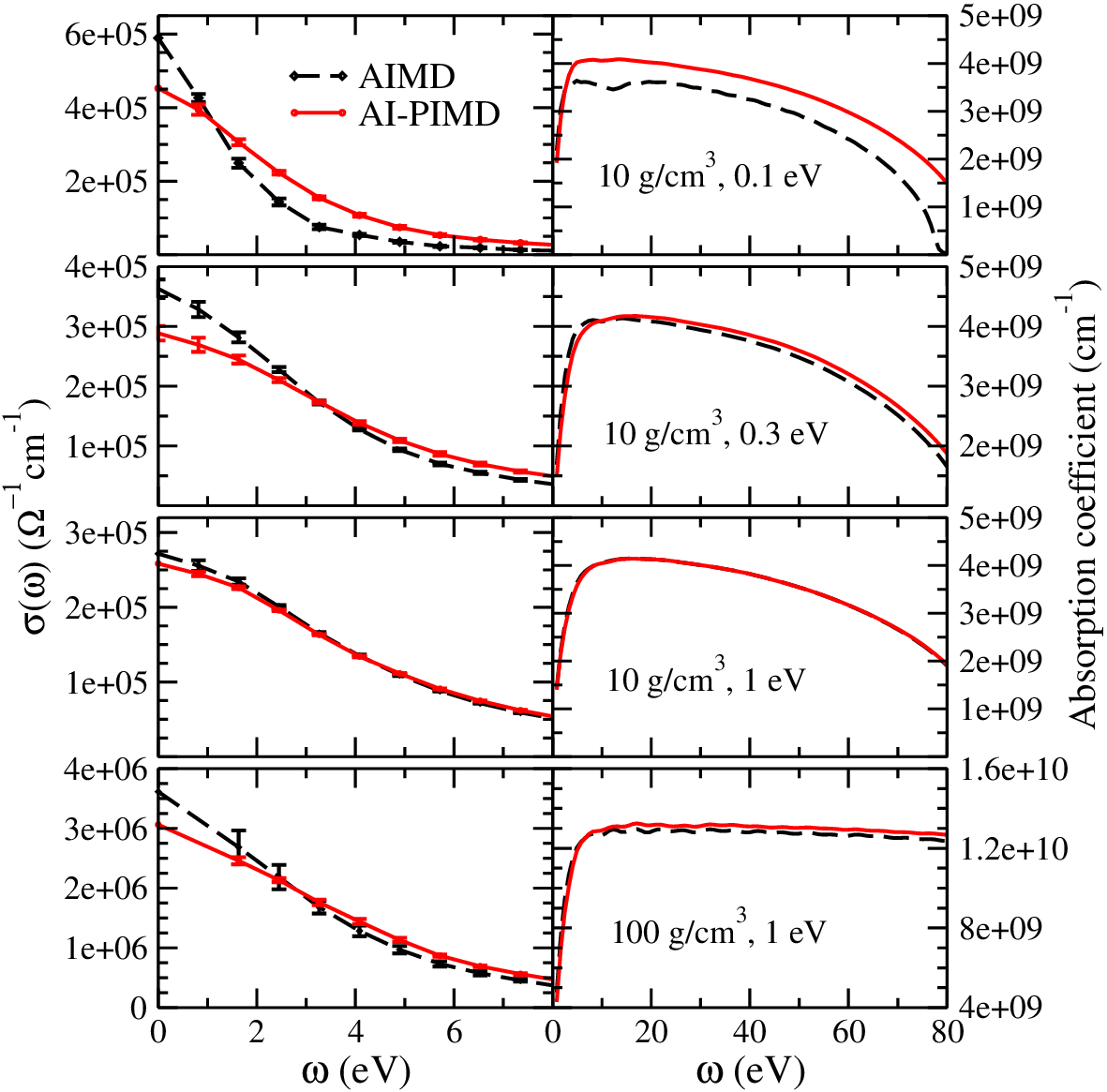}
\caption{(color online). Electrical conductivity and optical absorption coefficient from the AI-PIMD (solie lines) and AIMD (dashed lines) calculations for there state points, \ie, 10 g/cm$^3$ and 0.1 eV (upper panel), 10 g/cm$^3$ and 0.3 eV (second panel), 10 g/cm$^3$ and 1 eV (third panel), 100 g/cm$^3$ and 1 eV (lower panel). The error bars denote the standard deviation of averaging different atomic configurations.}
\label{fig3}
\end{figure}

Of our particular interest is whether the NQEs has evident effects on the electronic transport properties as profound as nuclei exhibits. To shed light on this issue, we calculated the electrical conductivity and optical absorption coefficient of dense hydrogen. In order to avoid the artificial drop at low frequency due to the finite number of atoms in the simulations, we employed the Drude formula to estimate the dc conductivity at zero frequency limit \cite{Collins}. It can be clearly seen from Fig.~\ref{fig3} that with the inclusion of NQEs, the low-frequency dependent electrical conductivity $\sigma(\omega)$ exhibits different trend compared with the results from the AIMD simulations, which directly leads to the different dc conductivity $\sigma_{dc}(\omega\rightarrow 0$ except for the state point of 10 g/cm$^3$ and 1 eV. The dc conductivity from the AI-PIMD calculations are largely suppressed compared with the AIMD value by 30\% at 10 g/cm$^3$, 0.1 eV, 25\% at 10 g/cm$^3$, 0.3 eV and 18\% at 100 g/cm$^3$, 1 eV. Interestingly, a visible enhancement of optical absorption for hydrogen of 10 g/cm$^3$ at 0.3 eV and 100 g/cm$^3$ at 1 eV, displayed in Fig.~\ref{fig3}. It is well known that NQEs will introduce more disorder for the systems, lowering the melting point and critical point \cite{Morales}. In fact, NQEs introduce ionic delocalization and closer ionic distances, as shown in Fig.~\ref{fig2}, which produce more localized electrons around the nuclei \cite{Supplement,Dai1}. The stronger localized electrons might enhance the optical oscillator strength and optical absorption cross sections of relative high frequencies. In addition, NQEs cause nonuniform broaden of energy bands because of the anharmonic effects, resulting significant difference in the electronic density of states (DOS). In particular, the DOSs around Fermi level decrease by NQEs \cite{Supplement}.

\begin{figure}[t]
\centering
\includegraphics*[width=8cm]{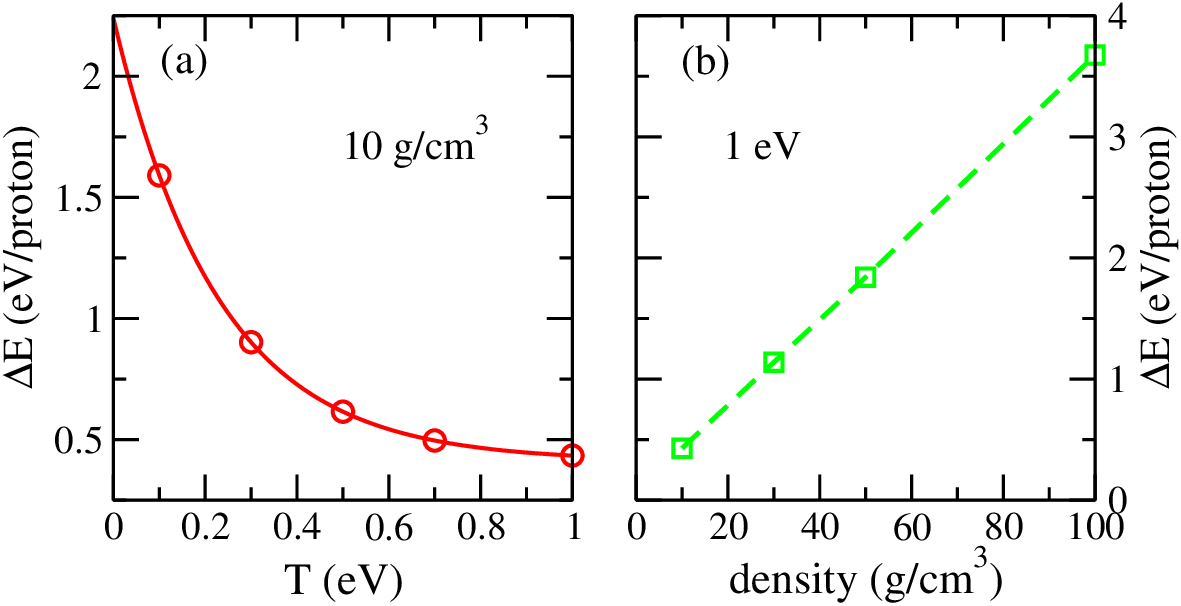}
\caption{(color online). Differences of total energy between the AI-PIMD and AIMD calculations at a density of 10 g/cm$^3$ (a) and a temperature of 1 eV (b).}
\label{fig4}
\end{figure}

To clarify NQEs on equation of state, we calculated the energies and pressures, and note that both the energies and pressures including NQEs are higher than the values of AIMD simulations over the whole temperature and density range in this work \cite{Supplement}. While the ZPM and anharmonic effects account for the increase of ionic kinetic energies and ionic pressures, the quantum spacial delocalization allows the nuclei to move closer and the Pauli exclusion principles enhances the kinetic energies of electrons that surround nuclei, thus the energies and pressures contributed by electrons rise inevitably. Figure~\ref{fig4} shows the differences of total energies between the AI-PIMD and AIMD calculations. The energy difference $\Delta E$ is dropped at an exponential rate with increasing temperature, while raised linearly with increasing density. We note that the zero-point energy (ZPE) is important for determining the stable high-pressure structure \cite{McMahon2,Liu,Takezawa} and melting point of dense hydrogen \cite{Liberatore}, but precisely estimating ZPE beyond harmonic approximation is still a challenge in dense hydrogen \cite{Takezawa}. The actual ZPE would be much larger than that estimated by harmonic treatment at high pressures due to the strong anharmonicity of nuclear motions \cite{Natoli}. Here we fitted the temperature-dependent energy differences shown in Fig.~\ref{fig4}(a) to an exponential expression and obtained the ZPE at 10 g/cm$^3$ is 2.24 eV/proton \cite{ZPE}, which is much larger than that we could obtain from extrapolating the ZPE curve in Ref. \onlinecite{McMahon2} (about 1.78 eV/proton). It is due to that our PIMD calculations include anharmonicity of the nuclear motions naturally. In fact, ZPM can also strongly affect the optical absorption at low temperatures, inducing large corrections to the optical and electronic properties \cite{ZPA}, which can explain partly the above differences of the electron and photon transport behaviors induced by NQEs.

In conclusion, we have performed AIMD and AI-PIMD simulations to study the structure and transport properties of dense liquid hydrogen. The complex transport behaviors show that even when the NQEs have little effect on the structures (10g/cm$^3$ and 1 eV), the ionic diffusions are also significantly enhanced due to the lower collision cross sections with the inclusion of NQEs. Electronic transport properties are less sensitive to NQEs here than ionic diffusions, but also seriously affected at relatively low temperature. The energy and pressure from the AI-PIMD simulations are higher than the AIMD value in the whole range of conditions here. The ZPE at 10 g/cm$^3$ for hydrogen is estimated by fitting the differences of total energies between the AI-PIMD and AIMD calculations at a temperature ranging from 0.1 to 1 eV. We can see that the quantum nuclear character will induce complex behaviors for both the ionic and electronic transports of dense hydrogen and can not be neglected even up to warm dense regime.

This work is supported by the National Natural Science Foundation of China under Grant Nos. 11274383, 11104351 and 11005153, the Major Research plan of National NSF of China (Grant No. 91121017). DDK thanks the support of the Hunan Provincial Innovation Foundation for Postgraduate under Grant No. CX2011B009. Calculations were carried out at the Research Center of Supercomputing Application, NUDT.


\begin{thebibliography}{99}
\bibitem{PP} After careful tests, a Coulombic pseudopotential with plane-wave cutoffs of 200-400 Ry was adopted, whose performance has been verified in previous studies in Ref. \onlinecite{Dai2}. The $\Gamma$ point was used for sampling the Brillouin zone when performing molecular dynamics simulations.
\bibitem{Timestep} Time step of 2-3 a.u. was used in our AI-PIMD calculations with 30000 steps after thermalization, while the smaller time steps of 0.2-0.3 a.u. with 10$^5$ steps and a centroid adiabaticity parameter of 20 were adopted in the AI-CMD simulations in order to decouple the centroid mode from normal modes.
\bibitem{ZPE} The exponential expression is $\Delta E=\textrm{a}+\textrm{b}\exp(-\textrm{c}T)$ (a, b and c are 0.411132, 1.82818, 4.3833, respectively). When $T = 0$, $\Delta E$ is the so-called zero-point energy.
\end{thebibliography}
\end{document}